\documentclass[aps,prd,twocolumn,showpacs,preprintnumbers,nofootinbib,amsmath,amssymb]{revtex4}

\usepackage{graphicx}
\usepackage{dcolumn}
\usepackage{bm}
\usepackage{amsmath}
\usepackage{braket}
\usepackage[hidelinks]{hyperref}
\usepackage{epstopdf}
\usepackage{placeins}
\usepackage{multirow}
\usepackage{makecell}
\usepackage{cleveref}

\newcommand{\beq}{\begin{eqnarray}}
\newcommand{\eeq}{\end{eqnarray}}
\newcommand{\beqnn}{\begin{eqnarray*}}
\newcommand{\eeqnn}{\end{eqnarray*}}

\newcommand{\CP}{\mathrm{CP}}
\newcommand{\SU}{\mathrm{SU}}
\newcommand{\cool}{\mathrm{cool}}

\usepackage[font=small, labelfont=bf, textfont={small,it}]{caption}
\def\spose#1{\hbox to 0pt{#1\hss}}
\def\ltapprox{\mathrel{\spose{\lower 3pt\hbox{$\mathchar"218$}}
 \raise 2.0pt\hbox{$\mathchar"13C$}}}

\begin{document}

\title{Lattice determination of the topological susceptibility slope $\chi^\prime$ of $2d~\mathrm{CP}^{N-1}$ models at large $N$}

\author{Claudio Bonanno}
\email{claudio.bonanno@fi.infn.it}
\affiliation{INFN Sezione di Firenze,\\ 
Via G.~Sansone 1, I-50019 Sesto Fiorentino, Firenze, Italy
}

\date{\today}

\begin{abstract}
We compute the topological susceptibility slope $\chi^\prime$, related to the second moment of the two-point correlator of the topological charge density, of $2d$ $\mathrm{CP}^{N-1}$ models for $N=5,11,21$ and $31$ from lattice Monte Carlo simulations. Our strategy consists in performing a double limit: first, we take the continuum limit of $\chi^\prime$ at fixed smoothing radius in physical units; then, we take the zero-smoothing-radius limit. Since the same strategy can also be applied to $4d$ gauge theories and full QCD, where $\chi^\prime$ plays an intriguing theoretical and phenomenological role, this work constitutes a step towards the lattice investigation of this quantity in such models.
\end{abstract}


\maketitle

\section{Introduction}\label{sec:intro}

The space of gauge field configurations with finite Euclidean action can be divided in disconnected sectors characterized by an integer number, the topological charge
\beq\label{eq:general_def_topocharge}
Q = \int d^d x \, q(x) \in \mathbb{Z},
\eeq
where $d$ is the space-time dimension and $q(x)$ is the topological charge density, which is a local function of the gauge fields. Many intriguing non-perturbative properties of $4d$ non-abelian gauge theories, as well as many phenomenological aspects of the Standard Model (and beyond), are related to the topology of gauge fields.

A particularly interesting topological quantity to consider is the $2$-point correlation function of $q(x)$~\cite{Alles:1997ae,Vicari:1999xx,Horvath:2005cv,Vicari:2008jw,Chowdhury:2012sq,Fukaya:2015ara,Mazur:2020hvt,Altenkort:2020axj,Mancha:2022umq}:
\beq\label{eq:def_2pnt_corr_q_fourier}
\widetilde{G}(p^2) \equiv \int d^d x \, e^{i p \cdot x} G(x),
\eeq
where $G(x)$ is the Euclidean space-time correlator
\beq\label{eq:def_2pnt_corr_q}
G(x-y) \equiv \braket{q(x) q(y)} = \braket{q(x-y) q(0)}.
\eeq

In Fourier space, the $2$-point correlator is a function of $p^2 \equiv p_\mu p_\mu$, being $q(x)$ a $\mathrm{CP}$-odd operator, and it is customary to Taylor-expand it around $p^2 = 0$ in powers of $p^2$:
\beq\label{eq:G_mom_exp}
\widetilde{G}(p^2) = \sum_{n=0}^{\infty} (-1)^{n} G_{2n} \, p^{2n},
\eeq
where the expansion coefficients $G_{2n}$ are proportional to the even moments of the correlator $G(x)$:
\beq\label{eq:G_2n_definition}
G_{2n} \equiv \frac{1}{d^n(2n)!} \int d^dx \vert x \vert^{2n} G(x).
\eeq

The leading order coefficient of the expansion in Eq.~\eqref{eq:G_mom_exp} is nothing but the integral of $G(x)$, which is equal to the well-known \emph{topological susceptibility}
\beq
\chi \equiv \underset{V\to\infty}{\lim} \frac{\braket{Q^2}}{V},
\eeq
as it can be easily derived from the translation-invariance of the correlator:
\begin{align*}
\widetilde{G}(0) = G_0 &= \int d^dx \braket{q(x)q(0)} \\
&= \underset{V\to\infty}{\lim} \frac{1}{V}\int d^d x\, d^d y \, \braket{q(x)q(y)} \\
&=\underset{V\to\infty}{\lim} \frac{\braket{Q^2}}{V} = \chi.
\end{align*}

The object of investigation of the present paper is, instead, the next-to-leading-order term in the momentum expansion of $\widetilde{G}(p^2)$, the so-called \emph{topological susceptibility slope} $\chi^\prime$:
\beq\label{eq:def_chi_prime_continuum}
\begin{aligned}
\chi^\prime &\equiv - \frac{d \widetilde{G}(p^2)}{dp^2} \Bigg\vert_{p^2=0}\\
&= \int d^d x \, \frac{\vert x \vert^2}{2d} \braket{q(x)q(0)}.
\end{aligned}
\eeq

This quantity is extremely interesting from a theoretical and from a phenomenological point of view. In $\SU(N)$ pure-gauge theories, the susceptibility slope controls the internal consistency of the Witten--Veneziano mechanism~\cite{Witten:1978bc,Veneziano:1979ec,DiVecchia:1980yfw}. More precisely, for the Witten--Veneziano equation for the $\eta^\prime$ mass to be valid, $\vert \chi^\prime \vert \ll \chi/m^2_{\eta^\prime}$ must hold in the large-$N$ limit. In QCD, instead, the value of the susceptibility slope in the chiral limit (which is expected to be non-vanishing, unlike the chiral limit of $\chi$) is related to the ``spin content'' of the proton via the so-called $\mathrm{U}(1)$ Goldberger--Treiman relation~\cite{Shore:1990zu,Shore:1992rg,Alles:1995aw,Narison:1998aq,Bernard:2001rs}.

However, in spite of its interesting theoretical and phenomenological role, the computation of $\chi^\prime$ has been quite disregarded for what concerns the lattice approach, especially if compared with the topological susceptibility, which has been widely investigated by numerical Monte Carlo (MC) simulations both in pure-gauge theories~\cite{Alles:1996nm, Alles:1997qe, DelDebbio:2004ns, DelDebbio:2002xa, DElia:2003zne,DelDebbio:2006yuf, Lucini:2004yh,Giusti:2007tu, Vicari:2008jw, Panagopoulos:2011rb, Ce:2015qha, Ce:2016awn,Bonati:2015sqt, Bonati:2016tvi, Bonati:2018rfg, Burger:2018fvb,Bonati:2019kmf,Athenodorou:2020ani,Bonanno:2020hht,Athenodorou:2021qvs,Bonanno:2022yjr,Bennett:2022ftz} and in full QCD~\cite{Bonati:2015vqz, Frison:2016vuc, Borsanyi:2016ksw, Petreczky:2016vrs, Bonati:2018blm,Burger:2018fvb,Bonanno:2019xhg,Lombardo:2020bvn, Kotov:2021ujj, Athenodorou:2022aay, Chen:2022fid}, but also in lower-dimensional toy models~\cite{Campostrini:1988cy,Campostrini:1992ar,Campostrini:1992it,Alles:1997nu,DelDebbio:2004xh,Bietenholz:2010xg,Hasenbusch:2017unr,Bonati:2017woi,Bonanno:2018xtd,Berni:2019bch,Berni:2020ebn,Bonanno:2022dru}.

As a matter of fact, while a prediction for $\chi^\prime$ is available in QCD from Chiral Perturbation Theory~\cite{Leutwyler:2000jg} and from the QCD Sum Rule~\cite{Ioffe:1998sa,Narison:1998aq,Narison:2006ws,Narison:2021svo} (which also provides an estimation of $\chi^\prime$ in the pure $\SU(3)$ gauge theory~\cite{Narison:1998aq,Narison:2006ws}), only preliminary attempts to compute it from the lattice can be retrieved in the literature, cf.~Refs.~\cite{DiGiacomo:1990ij,Briganti:1991pb,Digiacomo:1992jg,Boyd:1997nt}.

The goal of the present work is thus to fill a gap in the lattice literature by making progress in the numerical computation of $\chi^\prime$. To this end, we start our investigation from the simpler framework offered by large-$N$ $2d$ $\CP^{N-1}$ models~\cite{DAdda:1978vbw, Luscher:1978qe,Vicari:2008jw, Shifman:2012zz}, which are interesting quantum field theories that have been extensively used as a theoretical laboratory for QCD. On one hand, they are much cheaper to simulate on the lattice with Monte Carlo methods. On the other hand, they offer the possibility to obtain analytical results about topological quantities in the large-$N$ limit through the $1/N$ expansion, which can be compared to lattice determinations.

Since the methods that will be applied to these models to compute $\chi^\prime$ are rather general, and can be applied also to $\SU(N)$ pure Yang--Mills or to full QCD, the main purpose of this work is to show the feasibility and the solidity of our numerical strategies to obtain $\chi^\prime$ from lattice simulations. Therefore, this paper constitutes a step towards a deeper investigation of $\chi^\prime$ from the lattice also in $4d$ gauge theories.

This paper is organized as follows. In Sec.~\ref{sec:CPN_continuum} we briefly recall the continuum definition of $\CP^{N-1}$ models and the available analytic predictions for the $1/N$ expansion of $\chi^\prime$. In Sec.~\ref{sec:setup} we outline our lattice setup and MC algorithms, as well as our numerical strategies to compute $\chi^\prime$. In Sec.~\ref{sec:results} we show our results for $\chi^\prime$ for $N=5,11,21$ and $31$ and we compare them with large-$N$ analytic predictions. Finally, in Sec.~\ref{sec:final} we draw our conclusions and discuss future outlooks of this work.

\section{Continuum $2d$ $\CP^{N-1}$ models and the analytic $1/N$ expansion of $\chi^\prime$}\label{sec:CPN_continuum}
The continuum Euclidean action of $2d$ $\CP^{N-1}$ models can be written as~\cite{DAdda:1978vbw,Luscher:1978qe}:
\beq\label{eq:continuum_action}
S[z,A]=\int d^2x\left[ \frac{N}{g}\overline{D}_\mu \overline{z}(x) D_\mu z(x) \right],
\eeq
where the complex scalar field $z=(z_1,\dots,z_N)$ is a $N$-component vector with unit norm $\overline{z}z = 1$, and where $D_\mu \equiv \partial_\mu + i A_\mu$ is the ordinary $\mathrm{U}(1)$ covariant derivative. The field $A_\mu(x)$ is a non-propagating $\mathrm{U}(1)$ gauge field which can be integrated out and expressed in terms of $z$~\cite{Luscher:1978qe,Vicari:2008jw}, but this formulation is more convenient for the purpose of the lattice discretization. Finally, the coupling constant $g$ is the \emph{'t Hooft coupling}, and it is kept constant as $N$ is varied.

The integer-valued topological charge, instead, reads~\cite{DAdda:1978vbw,Luscher:1978qe}:
\beq\label{eq:continuum_topcharge}
Q=\int d^2x \, q(x)= \frac{1}{2\pi} \epsilon_{\mu\nu} \int d^2x \, \partial_\mu A_\nu(x) \in \mathbb{Z},
\eeq
where $q(x)$ is the topological charge density, which is used to define $\chi^\prime$ as in Eq.~\eqref{eq:def_chi_prime_continuum}.

In the large-$N$ limit, it is possible to perform an analytic computation of the susceptibility slope within the framework of the $1/N$ expansion. In Ref.~\cite{Campostrini:1991kv}, the Leading Order (LO) and the Next-to-Leading-Order (NLO) terms in the $1/N$ expansion were computed:
\beq\label{eq:chip_largeN_analytic}
\chi^\prime = -\frac{3}{10 \pi} \frac{1}{N} + e_2^\prime \frac{1}{N^2} + O\left(\frac{1}{N^3}\right),
\eeq
where the NLO coefficient is $e_2^\prime \simeq 1.53671$.

\section{Numerical setup}\label{sec:setup}

In this section we discuss our numerical setup by outlining the adopted discretizations for the main quantities of interest, the adopted MC algorithms and the strategies we followed to compute the topological susceptibility slope $\chi^\prime$.

\subsection{Lattice discretization}\label{sec:lat_discr}

We discretize~\eqref{eq:continuum_action} on a square periodic lattice (we will discuss other boundary conditions later) with $L^2$ sites and lattice spacing $a$ by means of the $O(a)$ tree-level Symanzik-improved lattice action~\cite{Campostrini:1992ar}:
\beq\label{eq:lattice_action}
\begin{aligned}
S_L = &-2N\beta_L\sum_{x,\mu} \left \{ c_1 \Re\left[\overline{U}_\mu(x)\bar{z}(x+\hat{\mu})z(x)\right] \right. \\
&\left. + c_2 \Re\left[\overline{U}_\mu(x+\hat{\mu})\overline{U}_\mu(x)\overline{z}(x+2\hat{\mu})z(x)\right]
 \right\},
\end{aligned}
\eeq
where $\beta_L$ is the lattice bare coupling, and where we introduced site variables $z(x)$ satisfying $\overline{z}(x)z(x)=1$ and $\mathrm{U}(1)$ link variables $U_\mu(x)$ satisfying $\overline{U}_\mu(x)U_\mu(x)=1$ as well (no sum on $\mu$ is implied). The choice of the improved lattice action allows to exactly cancel out logarithmic corrections to the leading-order $O(a^2)$ behavior of the finite-lattice-spacing corrections to the continuum limit, improving convergence.

Lattice configurations are updated by means of the standard Over-Relaxation (OR) and over-Heat-Bath (HB)~\cite{Campostrini:1992ar} local algorithms. In particular, our standard updating step is made of $4$ lattice sweeps of OR, followed by $1$ sweep of HB. In the following, we will refer to this combination as the ``standard algorithm''.

The continuum topological charge~\eqref{eq:continuum_topcharge} can be discretized by several different definitions. In this work we will consider the \emph{geometric} definition built in terms of the gauge links~\cite{Campostrini:1992ar}:
\beq\label{eq:geometric_lattice_charge}
Q_L = \sum_x q_L(x) = \frac{1}{2\pi}\sum_{x} \Im \left\{ \log \left[\Pi_{01}(x)\right] \right\},
\eeq
with $\Pi_{\mu\nu}(x) \equiv U_\mu(x)U_\nu(x+a\hat{\mu})\overline{U}_\nu(x+a\hat{\mu})\overline{U}_\nu(x)$ the plaquette, and $q_L(x)$ the lattice topological charge density. This definition has the property of always resulting in an integer number: $Q_L \in \mathbb{Z}$. However, being it affected by \emph{dislocations}~\cite{Berg:1981nw,Campostrini:1992it}, UV fluctuations at the scale of the lattice spacing that can make identifying the correct winding number ambiguous, it is customary to compute the lattice topological charge after smoothing, using one's favorite algorithm, in order to dump short-distance fluctuations. We postpone a more detailed discussion on smoothing methods to Sec.~\ref{sec:chip_cooling}, where we will cover this topic more thoroughly in relation with $\chi^\prime$.

In order to fix the physical scale, we introduce the \emph{second-moment correlation length}
\beq\label{eq:def_xi_continuum}
\xi^2 \equiv \frac{1}{\int G_P(x)d^2x}\int G_P(x) \frac{\vert x\vert^2}{4} d^2 x ,
\eeq
where $G_P(x)$ stands for the two-point connected correlator of $P_{ij}(x) \equiv z_i(x) \overline{z}_j(x)$: 
\beq\label{eq:projector_definition}
G_P(x) \equiv \braket{P_{ij}(x)P_{ij}(0)}-\frac{1}{N}.
\eeq
Introducing the dimensionless quantity $\xi_L = \xi/a$, it is possible to compute it from the lattice as~\cite{Caracciolo:1998gga}:
\beq\label{eq:def_xi_lattice}
\xi_L^2 = \frac{1}{4\sin^2\left(\pi/L\right)}\left[ \frac{\widetilde{G}_P^{(L)}(0,0)}{\widetilde{G}_P^{(L)}(2\pi/L,0)}-1 \right],
\eeq
where $\widetilde{G}_P^{(L)}(p_\mu)$ is the Fourier transform of $G_P^{(L)}(x)$, the straightforward lattice definition of $G_P(x)$.

In the continuum limit $a\to 0$, approached by taking $\beta_L \to \infty$, the correlation length expressed in lattice units diverges as $a$. Thus, it is possible to trade the vanishing lattice spacing limit for the $\xi_L \to \infty$ one. Therefore, finite-lattice-spacing corrections to the continuum expectation value of a certain observable $\braket{\mathcal{O}}_{\mathrm{cont}}$ can be expressed as:
\beq
\braket{\mathcal{O}}_{\mathrm{lat}}(\xi_L) = \braket{\mathcal{O}}_{\mathrm{cont}} + \frac{c}{\xi_L^2} + o\left(\frac{1}{\xi_L^2}\right).
\eeq

\subsection{Topological freezing and parallel tempering on boundary conditions}

At large $N$ and close to the continuum limit, it is well known that standard updating algorithms suffer for a severe \emph{topological critical slowing down}~~\cite{Lucini:2004yh,DelDebbio:2004xh, DelDebbio:2006yuf, Bonati:2016tvi, Bonanno:2018xtd,Bonanno:2019xhg,Bonanno:2020hht,Bonanno:2022yjr}. This means that the auto-correlation time of the topological charge, i.e., the number of updating steps needed to generate two field configurations with uncorrelated values of $Q$, rapidly grows when $\xi_L$ and/or $N$ is large. Numerical evidence suggests that this growth is exponential in $N$ and $\xi_L$~\cite{DelDebbio:2004xh, DelDebbio:2006yuf, Bonati:2016tvi, Bonanno:2018xtd}. This in practice results in the so-called \emph{topological freezing}, i.e., the Monte Carlo history of the topological charge gets stuck in a fixed topological sector and eventually no topological fluctuation is observed, requiring unfeasible long simulations to achieve a proper sampling of the path integral.

To deal with this computational problem we adopted the \emph{Parallel Tempering on Boundary Conditions} (PTBC) algorithm to mitigate the effects of topological freezing. Originally proposed by M.~Hasenbusch~\cite{Hasenbusch:2017unr} for $2d$ $\CP^{N-1}$ models (see also~\cite{Berni:2019bch}), and recently implemented also for $4d$ pure-gauge Yang--Mills theories~\cite{Bonanno:2020hht,Bonanno:2022yjr}, this algorithm has proven to be extremely effective in mitigating the effects of topological freezing at large $N$ by reducing the auto-correlation time of $Q$ by up to several orders of magnitude. Here, we use the same algorithmic setup of Ref.~\cite{Berni:2019bch}.

\begin{figure}[!htb]
\centering
\includegraphics[scale=0.49]{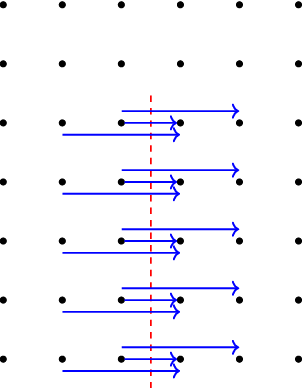}
\caption{Figure taken from Ref.~\cite{Berni:2019bch}. Dashed line represents the defect $D$, solid arrows represent links or product of links appearing in the Symanzik action~\eqref{eq:lattice_action} which orthogonally cross the defect line and get suppressed as factors of $c(r)$.}
\label{fig:defect}
\end{figure}

We consider $N_r$ copies of the discretized theory outlined in Sec.~\ref{sec:lat_discr}. Each replica differs from the others only for the boundary conditions imposed on a small sub-region of the lattice $D$, called \emph{the defect}. Away from the defect, boundary conditions are periodic as usual. In practice, we rewrite the lattice action~\eqref{eq:lattice_action} as:
\begin{multline*}\label{eq:lattice_action_replicas}
S_L^{(r)} = -2 N \beta_L\sum_{x,\mu}\left\{ k_\mu^{(r)}(x) c_1\Re\left[\bar{U}_\mu(x)\bar{z}(x+\hat{\mu})z(x)\right] \right. \\
\left.
+k_\mu^{(r)}(x+\hat{\mu})k_\mu^{(r)}(x)c_2\Re\left[\bar{U}_\mu(x+\hat{\mu})\bar{U}_\mu(x)\bar{z}(x+2\hat{\mu})z(x)\right]\right\} \, ,
\end{multline*}
where the coefficients
\beq
k_\mu^{(r)}(x) \equiv
\begin{cases}
c(r)\, , &\quad x \in D \wedge \mu=0\, ; \\
1\, , &\quad \text{otherwise.}\\
\end{cases}
\eeq
are used to effectively implement the chosen boundary conditions on the defect. In particular, the coefficients $c(r)$ are chosen so as to interpolate among periodic boundaries $c(0)=1$ and open boundaries $c(N_r-1)=0$. In our setup, we chose a simple linear interpolation: $c(r)=1-r/(N_r-1)$. The defect $D$ is set on the $\mu=1$ boundary and is $L_d$ sites long. A pictorial representation of the defect and of the sites/links affected by its presence is depicted in Fig.~\ref{fig:defect}.

Each replica is evolved independently using the standard algorithm earlier outlined. After the updates, field configurations are swapped among adjacent copies $(r,r+1)$, so that the field configuration can diffuse from the periodic replica to the open one and vice-versa, in a random-walk fashion. This way, the fast decorrelation of the topological charge achieved in the replica with open boundaries is transferred to the periodic copy, resulting in a strong reduction of the autocorrelation time of $Q$. The advantage of parallel tempering is that any physical observable can be computed on the periodic replica, thus avoiding technical complications related to finite size effects when computing correlators in open lattices.

The efficiency of the algorithm can be further improved by translating the periodic replica after the updates and the swaps, and by alternating updating sweeps over the full lattice with updating sweeps over small sub-lattices centered around the defect. These two ingredients allow, respectively, to move the position where new topological excitations are created/annihilated, and to increase the number of topological fluctuations that are created/annihilated (since they are mostly located in the neighborhood of the defect).

The number of replicas $N_r$ and the size of the defect $L_d$ are calibrated through short test runs in order to ensure that no swap acceptance $p(r,r+1)$ is smaller than $\sim 30\%$. In all cases, this was sufficient to ensure that a single field configuration uniformly explores all boundary conditions and does a random walk among the two extremes of the replicas chain.

\subsection{Computation of $\chi^\prime$ from the lattice}\label{sec:chip_cooling}

As mentioned in the previous section, to properly define the physical topological background of a lattice configuration, smoothing methods such as cooling~\cite{Berg:1981nw,Iwasaki:1983bv,Itoh:1984pr,Teper:1985rb,Ilgenfritz:1985dz,Campostrini:1989dh,Alles:2000sc}, stout smearing~\cite{APE:1987ehd, Morningstar:2003gk} or gradient flow~\cite{Luscher:2009eq, Luscher:2010iy} need to be employed to damp UV fluctuations. All choices give consistent results when
properly matched to each other~\cite{Alles:2000sc, Bonati:2014tqa, Alexandrou:2015yba}.

Such algorithms iteratively bring a configuration closer to a local minimum of the lattice action and smooth gauge fields up to a certain distance known as the \emph{smoothing radius} $r_s$, which is proportional to the square root of the amount of smoothing performed~\cite{Luscher:2009eq, Luscher:2010iy}. Since this procedure only dumps UV fluctuations at length scales $\lesssim r_s$, smoothing methods are expected to leave the global topological content of the field configuration unaltered.

As a matter of fact, lattice gluonic definitions of the topological charge like~\eqref{eq:geometric_lattice_charge} reach a plateau after a certain amount of smoothing. Therefore, the choice of the number of smoothing steps is not critical when one is interested in, e.g., the global topological charge $Q$ or the topological susceptibility $\chi \propto \braket{Q^2}$.

The situation is different when dealing with objects like the topological charge density correlator~\eqref{eq:def_2pnt_corr_q}. Indeed, if on one hand smoothing is necessary to correctly identify the true topological background of a configuration, on the other hand smoothing unavoidably modifies the short-distance behavior of $G(x)$, which is only properly recovered in the zero-smoothing limit. Therefore, to reliably compute topological quantities such as $\chi^\prime$, it is of the utmost importance to ensure that no relevant physical contribution coming from short distance fluctuations is lost due to the smoothing procedure.

For this reason, to extract the physical value of $\chi^\prime$ from lattice field configurations, we follow the same strategy applied also in Ref.~\cite{Altenkort:2020axj} to compute the topological charge density correlator: first, we perform the continuum limit fixing the smoothing radius $r_s$ in physical units; then, we extrapolate our results for finite $r_s$ towards the zero-smoothing limit $r_s \to 0$. 

In a $2d$ theory, $\chi^\prime$ is a dimensionless quantity that can be discretized on a periodic lattice as:
\beq 
\chi^\prime_L \equiv \frac{1}{4} \left\langle\sum_x d^2(x,0) q_L(x) q_L(0)\right\rangle,
\eeq
where $q_L(x)$ is the lattice topological charge density in Eq.~\eqref{eq:geometric_lattice_charge} and $d(x,y)$ is the shortest distance between lattice sites $x=(x_0,x_1)$ and $y=(y_0,y_1)$,
\begin{equation}
\begin{gathered}
d^2(x,y) = \sum_{\mu=0,1} d_\mu^2(x,y),\\
d^2_\mu(x,y) = 
\begin{cases}
(x_\mu - y_\mu)^2         ,& \vert x_\mu - y_\mu \vert < L/2, \\
[ L - (x_\mu - y_\mu) ]^2 ,& \vert x_\mu - y_\mu \vert > L/2.
\end{cases}
\end{gathered}
\end{equation}

We compute $\chi_L^\prime$ on the periodic replica after smoothing the field configuration. To this end, we adopt cooling for its numerical cheapness. A single cooling step is achieved aligning, site per site and link by link, each variable $z(x)$ and $U_\mu(x)$ to their relative local force. Since the action that is used to compute local forces during cooling needs not to be the same used for the Monte Carlo sampling~\cite{Alexandrou:2015yba}, we adopt the non-improved action to this end (i.e., action~\eqref{eq:lattice_action} with $c_1=1$ and $c_2=0$.)

To extrapolate finite-$\xi_L$ results of $\chi^\prime_L$ towards the continuum limit at fixed smoothing radius, it is sufficient to consider determinations obtained for the same value of
\beq
\frac{n_{\cool}}{\xi^2_L} \propto \left(\frac{r_s}{\xi}\right)^2,
\eeq
since the smoothing radius $r_s \propto a\sqrt{n_\cool}$:
\beq\label{eq:continuum_limit}
\chi^\prime_L\left(\xi_L, \frac{n_\cool}{\xi_L^2}\right) = \chi^\prime\left(\frac{n_\cool}{\xi_L^2}\right) + c \, \xi_L^{-2} + o\left(\xi_L^{-2}\right)
\eeq
where the coefficient $c$, in principle, may depend on the value of $n_\cool/\xi_L^2$.

Finally, we extrapolate the continuum extrapolations $\chi^\prime\left(n_\cool/\xi_L^2\right)$ appearing in the r.h.s.~of Eq.~\eqref{eq:continuum_limit} towards the \emph{zero-cooling} limit according to the law:
\beq\label{eq:zero_cool_limit_law}
\chi^\prime\left(\frac{n_\cool}{\xi_L^2}\right) &=& \chi^\prime + k \, \frac{n_\cool}{\xi_L^2} + o\left(\frac{n_\cool}{\xi_L^2}\right),
\eeq
where $\chi^\prime$ in the r.h.s.~of Eq.~\eqref{eq:zero_cool_limit_law} represents our final result for the susceptibility slope.

To justify Eq.~\eqref{eq:zero_cool_limit_law}, we can rely on the argument given in Ref.~\cite{Altenkort:2020axj} for the Wilson flow. Within the gradient flow framework, it is possible to express expectation values of flowed operators in terms of expectation values of unflowed operators via the Operator Product Expansion (OPE). In the OPE series, the contribution of higher-dimensional operators is compensated by suitable powers of the flow time $\tau_{\mathrm{flow}}$. In particular, when considering $q(x)q(0)$, the first correction to the zero-flow result is expected to be linear in $\tau_{\mathrm{flow}}$~\cite{Altenkort:2020axj}. Since it has been established the numerical equivalence between gradient flow and cooling, and the existence of the linear proportionality $n_\cool \propto \tau_{\mathrm{flow}}$\footnote{For example, in the $4d$ $\SU(3)$ pure-gauge theory, the correspondence $n_\cool = 3\tau_{\mathrm{flow}}$ holds adopting the Wilson gauge action~\cite{Bonati:2014tqa}.}~\cite{Bonati:2014tqa}, we expect to observe a linear dependence of $\chi^\prime(n_\cool/\xi_L^2)$ on $n_\cool/\xi_L^2$ (see also Ref.~\cite{Bonanno:2022dru} for a similar discussion concerning the linear zero-cooling extrapolation of $\chi(n_\cool)$ for the $2d$ $\CP^1$ model).

\section{Results}\label{sec:results}

\begin{table}[!htb]
\begin{center}
\begin{tabular}{|c|c|c|c|c|c|c|c|c|}
\hline
$N$ & $\beta_L$ & $L$ & $\xi_L$ & $L/\xi_L$ & $N_r$& $L_d$ & \makecell{Max num.\\cool.~steps} & \makecell{Stat.\\(M)} \\
\hline
\multirow{3}{*}{$5$}  & 1.00  & 200 & 13.419(37) & 14.9 &    &   & 110 & 17.5 \\
                      & 1.05  & 300 & 18.08(12)  & 16.6 &    &   & 200 & 6.9  \\
                      & 1.10  & 400 & 24.65(21)  & 16.2 &    &   & 370 & 4.7  \\
\hline
\multirow{4}{*}{$11$} & 0.75  & 105 & 5.499(27)  & 19.1 &    &   & 120 & 3.5 \\
                      & 0.77  & 140 & 6.369(67)  & 22.0 &    &   & 155 & 7.5 \\
                      & 0.79  & 140 & 6.987(67)  & 20.0 &    &   & 195 & 7.5 \\
                      & 0.848 & 220 & 9.90(12)   & 22.2 &    &   & 405 & 6.7 \\
\hline
\multirow{5}{*}{$21$} & 0.66  & 88  & 4.2129(78) & 20.9 & 9  & 5 & 250 & 10.7 \\
                      & 0.68  & 102 & 4.7561(99) & 21.4 & 10 & 6 & 320 & 10.7 \\
                      & 0.70  & 114 & 5.422(15)  & 21.0 & 11 & 6 & 415 & 5.9  \\
                      & 0.718 & 128 & 6.030(13)  & 21.2 & 11 & 6 & 515 & 9.5  \\
                      & 0.741 & 150 & 6.979(11)  & 21.4 & 11 & 6 & 690 & 22.4 \\
\hline
\multirow{4}{*}{$31$} & 0.58  & 72  & 2.8518(38) & 25.2 & 10 & 4 & 110 & 24.0 \\
                      & 0.60  & 82  & 3.2409(75) & 25.3 & 10 & 4 & 140 & 8.7  \\
                      & 0.62  & 92  & 3.6731(52) & 25.0 & 10 & 4 & 180 & 7.2  \\
                      & 0.715 & 172 & 6.585(37)  & 26.1 & 13 & 7 & 585 & 2.3  \\
\hline
\end{tabular}
\end{center}
\caption{Summary of simulation parameters. The PTBC algorithm was employed for $N=21,31$, while $N=5,11$ were simulated with the standard algorithm. Measures were taken every 10 updating steps for the standard algorithm, and at each step for parallel tempering runs. Total collected statistics is expressed in millions (M).}
\label{tab:simulation_summary}
\end{table}

In Tab.~\ref{tab:simulation_summary} we summarize all performed simulations, along with the total collected statistics and the maximum number of cooling steps computed for each simulation point.

\begin{figure}[!htb]
\centering
\includegraphics[scale=0.35]{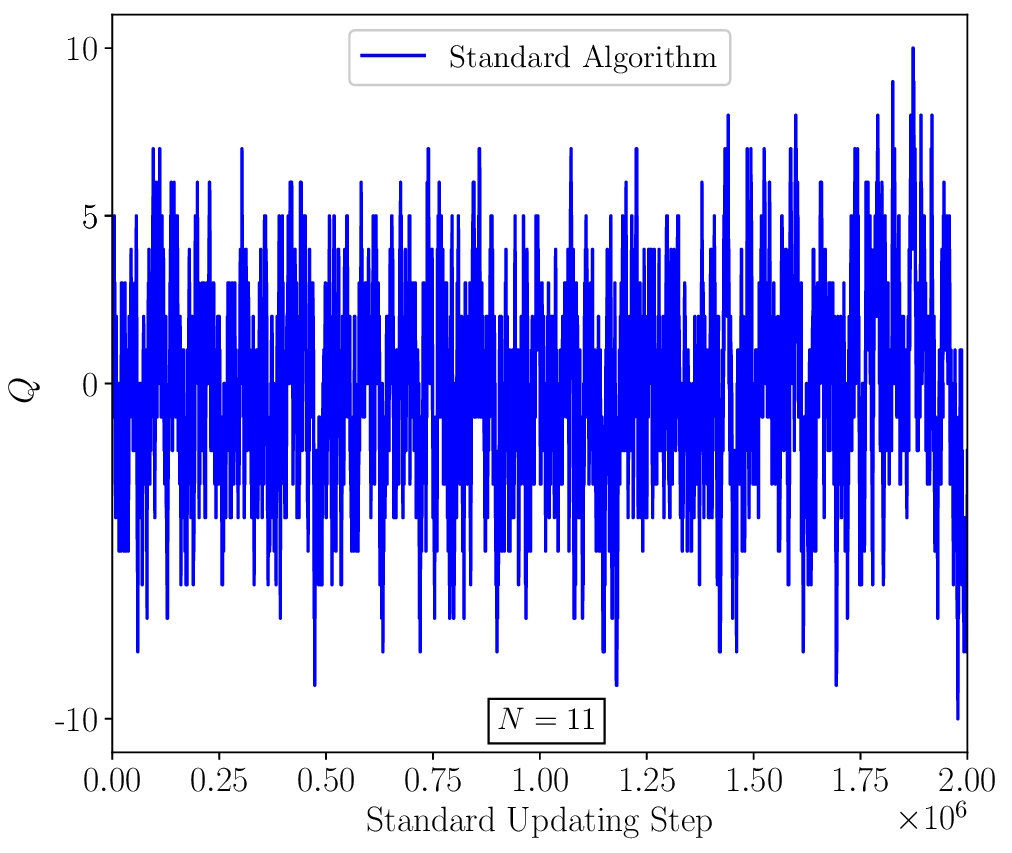}
\includegraphics[scale=0.35]{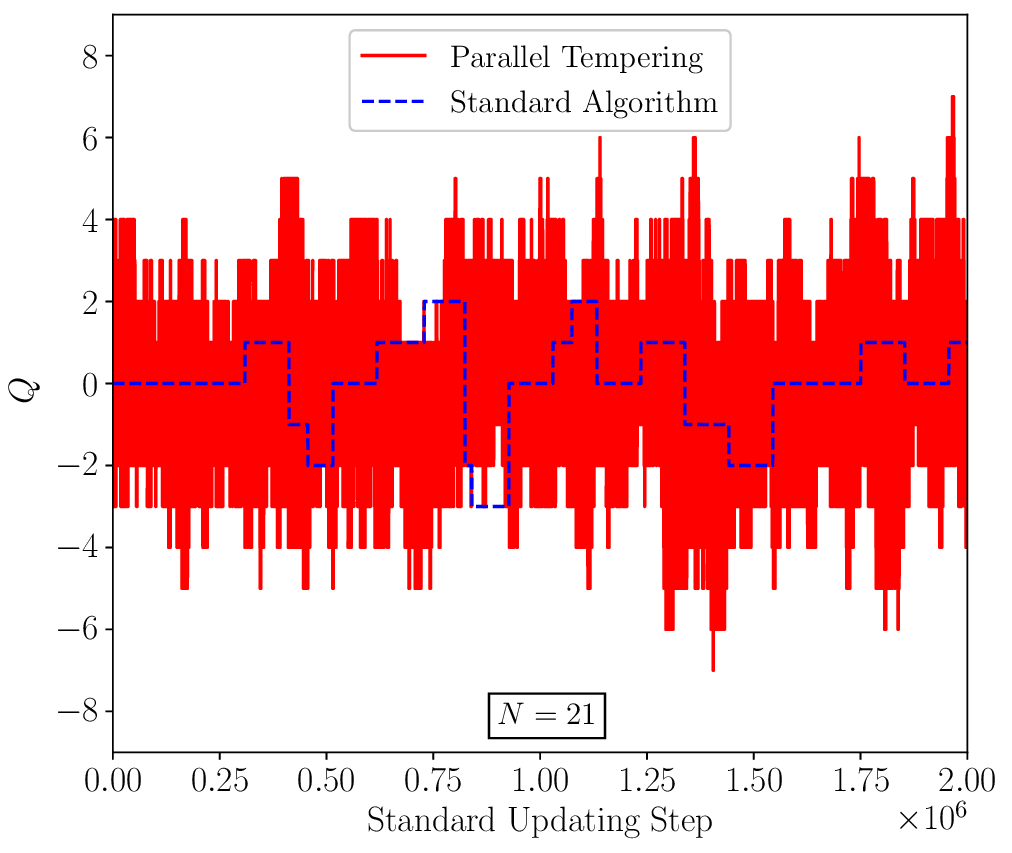}
\caption{Monte Carlo history of the topological charge $Q$. Plot refers to $N=11$, $\beta_L=0.848$ (top) and $N=21$, $\beta_L=0.741$ (bottom). Monte Carlo time for the PTBC algorithm has been expressed in units of standard updates by scaling for the number of replicas. Shown time windows correspond to, respectively, $\sim 14\%$ and $\sim 10\%$ of the total collected statistics.}
\label{fig:story_Q}
\end{figure}

Simulations for $N=21$ and $31$ where carried on adopting the PTBC algorithm, while for $N=5$ and $11$ standard updating algorithms were sufficient to easily decorrelate the topological charge. Illustrative examples are shown in Fig.~\ref{fig:story_Q}. In the top plot we show the Monte Carlo history of the topological charge $Q$ for $N=11$ and $\beta_L=0.848$. The integer charge $Q$ is obtained computing the geometric definition $Q_L$ in Eq.~\eqref{eq:geometric_lattice_charge} after $30$ cooling steps. In this case, the standard algorithm allows a uniform exploration of different topological sectors. In the bottom plot we show a comparison between the Monte Carlo evolutions of $Q$ obtained with the PTBC algorithm and the standard one for $\beta_L=0.741$ at $N=21$. As it can be appreciated, while the standard algorithm suffers for much longer auto-correlation times, the PTBC algorithm allows much more fluctuations of the topological charge during the same Monte Carlo time.

\begin{figure}[!htb]
\centering
\includegraphics[scale=0.38]{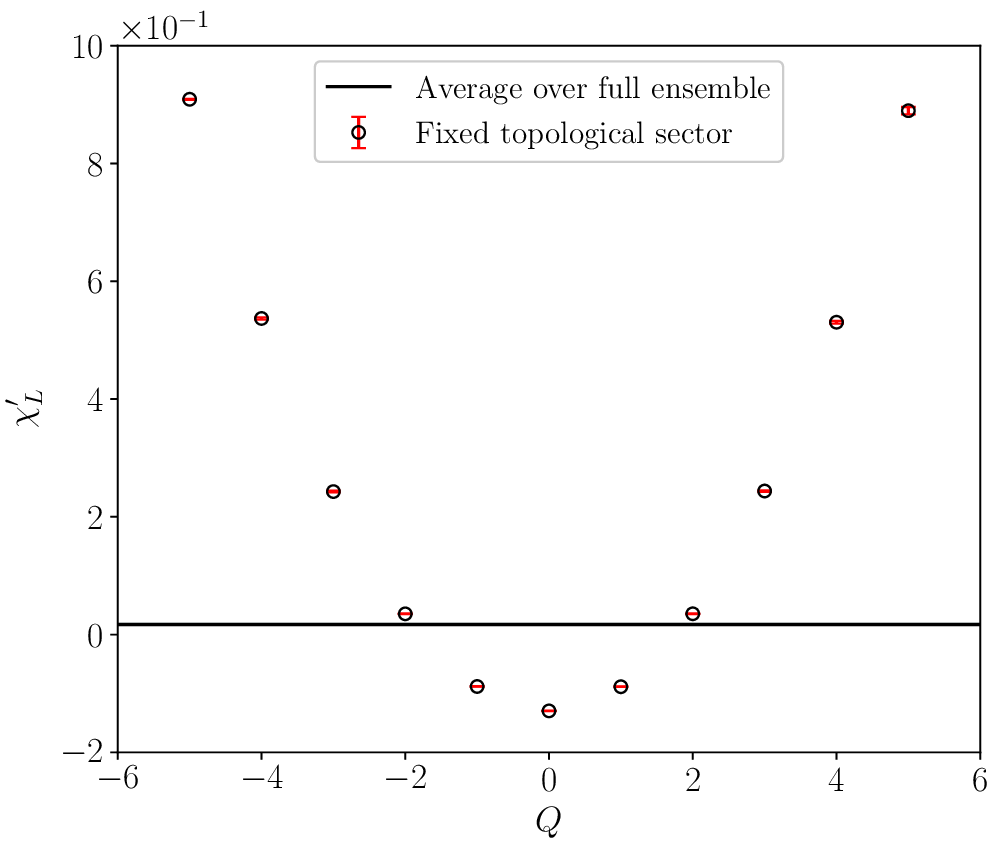}
\caption{Results for $\chi^\prime_L$ computed on sub-ensembles with fixed topological charge $Q$ as a function of the topological background. Horizontal line displays the result obtained averaging over the full ensemble. Error bars are not visible at this scale. Plot refers to $N=21$, $\beta_L=0.741$, $n_\cool=200$.}
\label{fig:chip_vs_Q}
\end{figure}

This aspect is crucial to correctly compute $\chi_L^\prime$. As a matter of fact, although the topological charge density fluctuates even in the $Q=0$ sector, the values obtained for $\chi_L^\prime$ restricting to sub-ensembles with fixed topological charge are quite different from the one obtained by taking the mean over the full ensemble, as it can be observed from Fig.~\ref{fig:chip_vs_Q}.

\begin{figure}[!htb]
\includegraphics[scale=0.38]{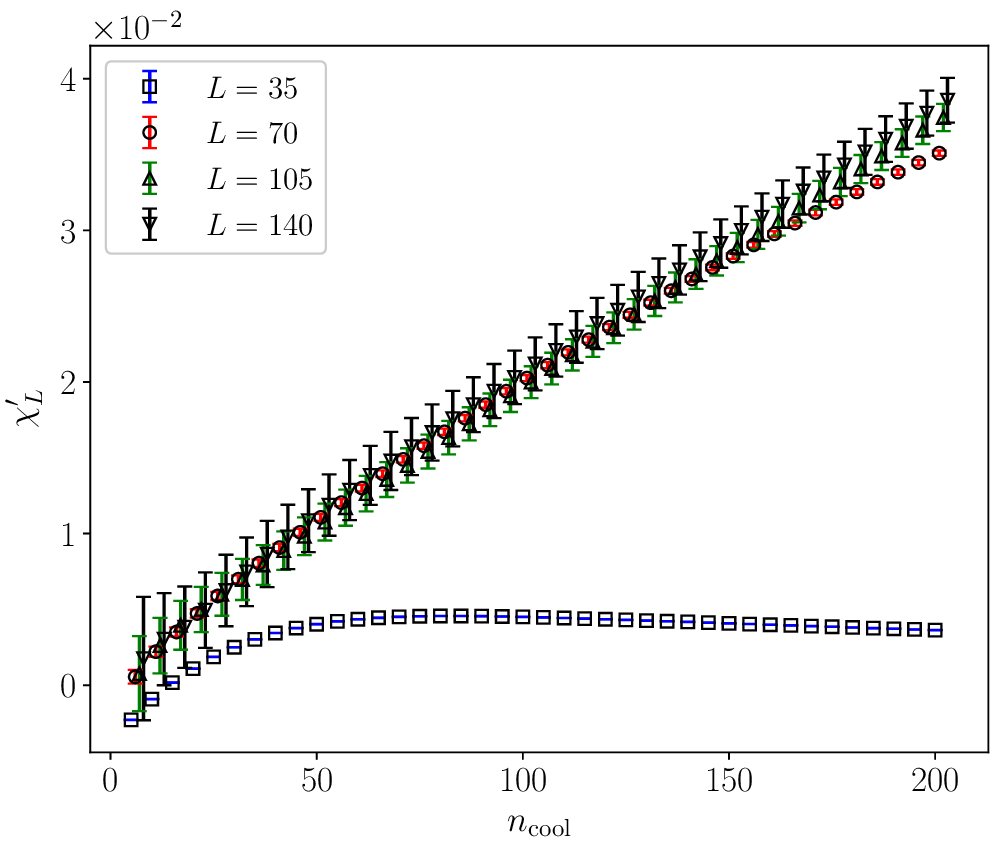}
\caption{Dependence of $\chi^\prime_L$ on the lattice size $L$ for several values of $n_\cool$. Plot refers to $N=11$ and $\beta_L=0.77$. Points referring to the same value of $n_\cool$ have been slightly shifted.}
\label{fig:finite_size_scaling_ex}
\end{figure}

For what concerns the magnitude of finite-size effects, in the large-$N$ limit two constraints have two be satisfied in order to have them under control: $L/\xi_L \gg 1$ and $(L/\xi_L)^2/N \gg 1$, as discussed in Ref.~\cite{Aguado:2010ex} (see also~\cite{Berni:2019bch} for a related discussion about this condition). In practice, it is sufficient to ensure $ (L/\xi_L)^2/N \gtrsim 20$ to have no significant systematic error related to finite volume~\cite{Bonanno:2018xtd,Berni:2019bch}. Satisfying this condition required $L/\xi_L \gtrsim 15-25$ for the values of $N$ we explored. We show an illustrative example of the magnitude of finite size effects on $\chi^\prime_L$ for $N=11$ in Fig.~\ref{fig:finite_size_scaling_ex}.

\subsection{Double extrapolation - Example for $N=21$}\label{sec:double_extrapolation_ex}

In this section we exemplify, for $N=21$, the procedure followed to achieve the double extrapolation of $\chi^\prime$ towards the continuum and the zero-cooling limit.

\begin{figure}[!t]
\centering
\includegraphics[scale=0.4]{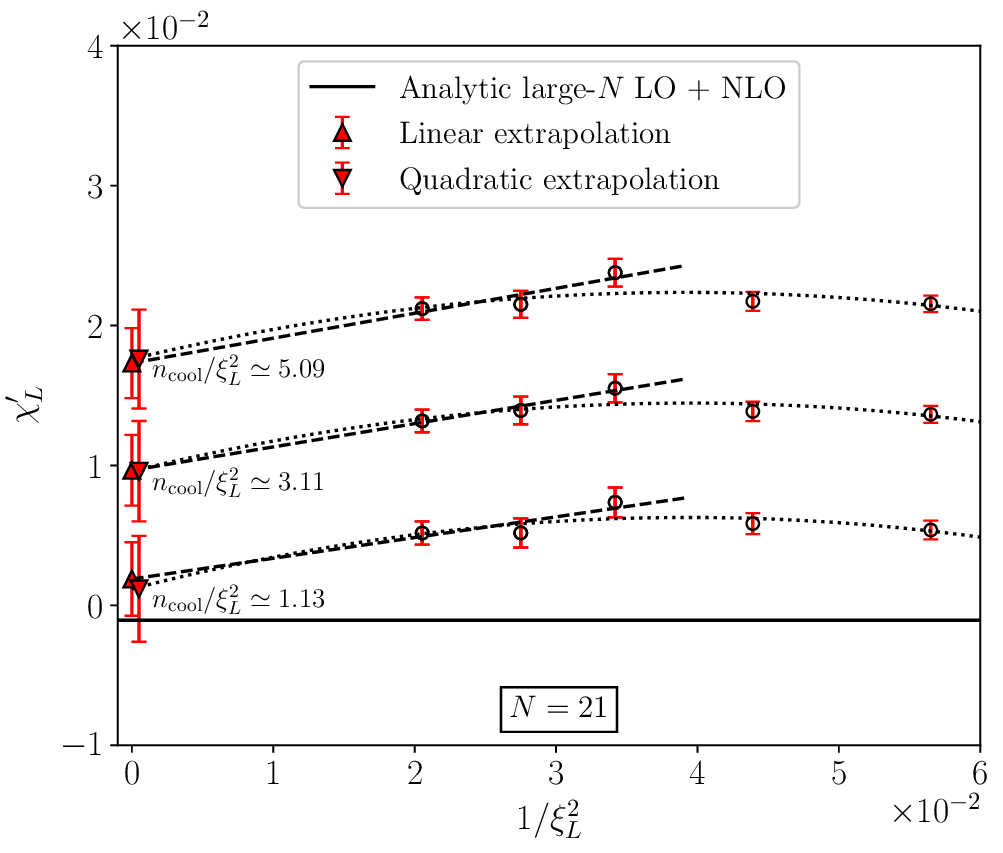}
\caption{Examples of extrapolation towards the continuum limit $\xi_L\to\infty$ of $\chi_L$ at fixed smoothing radius for $N=21$ and for $3$ values of $n_{\cool}/\xi_L^2$.}
\label{fig:cont_limit_ex}
\end{figure}

The first step is to extrapolate our data for $\chi^\prime_L$ computed at fixed value of $n_\cool / \xi_L^2$ towards the continuum limit according to Eq.~\eqref{eq:continuum_limit}. We performed a linear fit in $1/\xi_L^2$ considering the three largest correlation lengths available, but we also checked that performing the best fit in the whole available range gave exactly the same results within errors if further $O(1/\xi_L^4)$ corrections are taken into account, cf.~Fig.~\ref{fig:cont_limit_ex}. We thus conclude that a linear fit in $1/\xi_L^2$ to the 3 finest lattice spacing gives a solid continuum extrapolation. As a side note, we observe that no significant dependence on $n_\cool/\xi_L^2$ of the magnitude of corrections to the continuum limit is observed.

\begin{figure}[!t]
\centering
\includegraphics[scale=0.38]{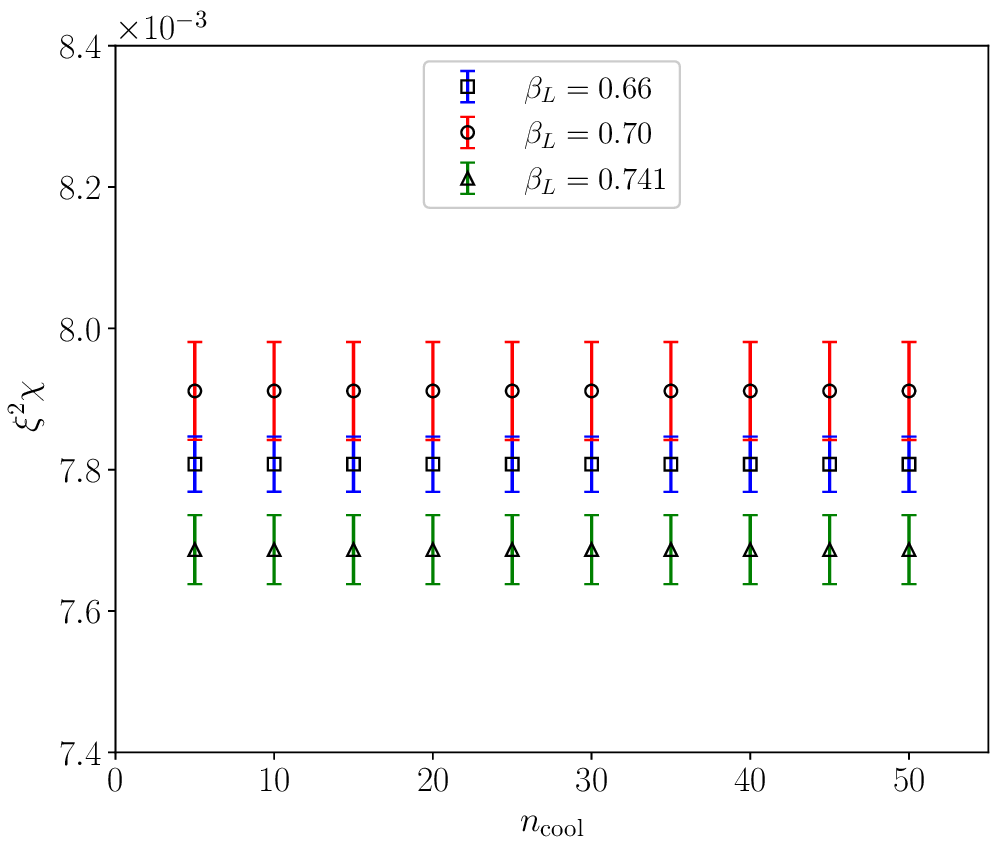}
\caption{Behavior of the topological susceptibility $\xi^2 \chi$ as a function of $n_\cool$ for three values of $\beta_L$ explored at $N=21$ (the largest, the smallest and the middle one).}
\label{fig:topsusc_vs_ncool_ex}
\end{figure}

\begin{figure}[!t]
\centering
\includegraphics[scale=0.38]{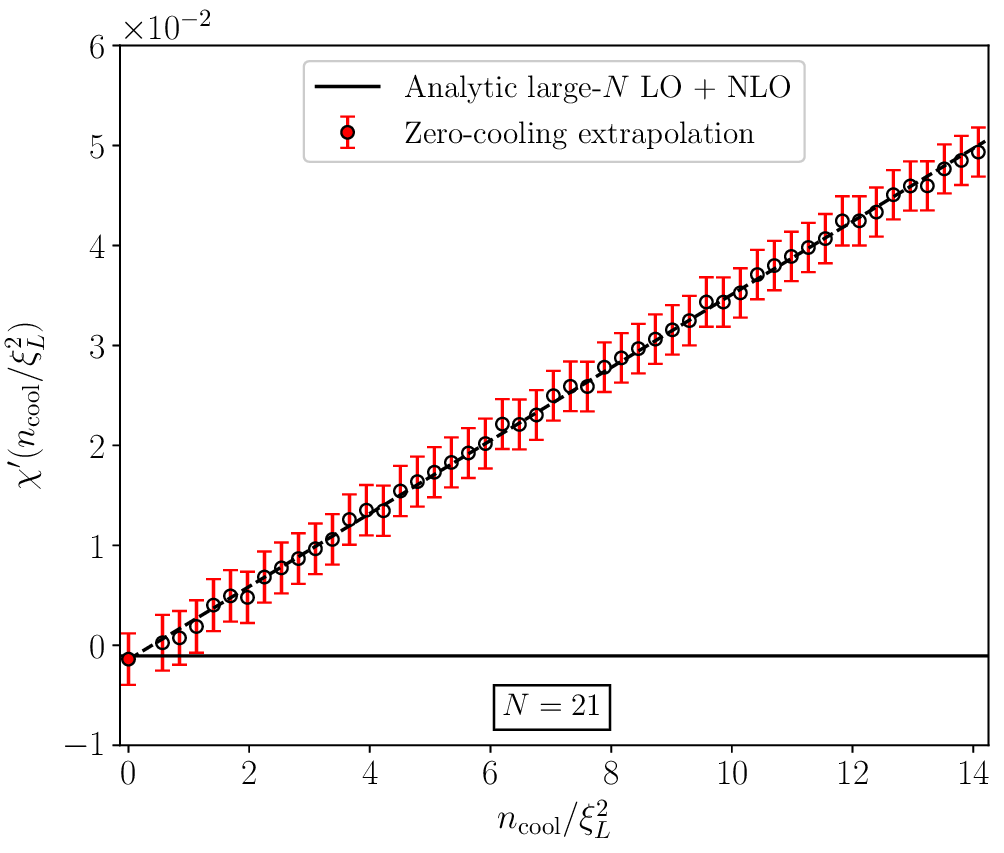}
\caption{Extrapolation towards the zero-cooling limit $n_\cool/\xi_L^2 \to 0$ of $\chi^\prime(n_\cool/\xi_L^2)$ for $N=21$ with a linear function in $n_\cool/\xi_L^2$. Full point represents our final result.}
\label{fig:zerocool_limit_ex}
\end{figure}

Once the continuum limit is taken, we can extrapolate our continuum results towards the zero-cooling limit by performing a linear best fit in $n_\cool/\xi_L^2$, according to Eq.~\eqref{eq:zero_cool_limit_law}. As it can be seen from Fig.~\ref{fig:zerocool_limit_ex}, our $\chi^\prime$ continuum extrapolations at finite smoothing radius are perfectly described by a linear law in $n_\cool/\xi^2_L \propto (r_s/\xi)^2$, in agreement with our expectations. For the purpose of comparison, in Fig.~\ref{fig:topsusc_vs_ncool_ex} we also report the LO $+$ NLO large-$N$ analytic prediction of Eq.~\eqref{eq:chip_largeN_analytic}.

Concerning the choice of the fit range in this case, the lower bound was fixed to $n_\cool/\xi_L^2 \sim 0.56$, corresponding to $n_\cool = 10$ for the smallest explored correlation length. As a matter of fact, we observed that $n_\cool \ge 10$ is enough to observe a plateau in the determinations of the topological susceptibility at finite $\xi_L$ as a function of $n_\cool$ for all simulated points, cf.~Fig.~\ref{fig:topsusc_vs_ncool_ex}. Therefore, this choice ensures that we are correctly identifying the topological background of the configurations for all the values of $n_\cool$ and $\xi_L$ employed in our analysis.

\begin{figure}[!t]
\centering
\includegraphics[scale=0.38]{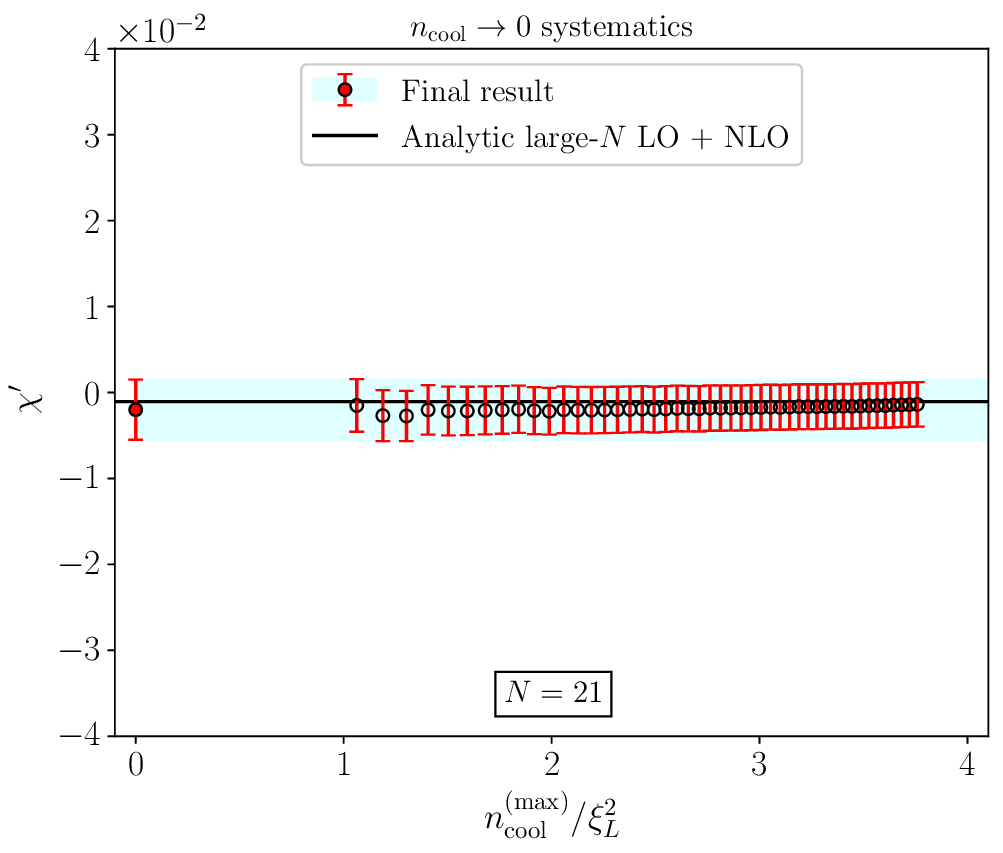}
\caption{Results of the zero-cooling extrapolation varying the upper bound of the fit range $n_\cool^{(\max)}/\xi_L^2$ for $N=21$, obtained by means of a bootstrap analysis. The lower bound was instead kept fixed at $n_\cool / \xi_L^2 \simeq 0.56$. The shaded area and the full point in $n_\cool/\xi_L^2=0$ represent our final result.}
\label{fig:systematics_zerocool}
\end{figure}

As for the upper bound of the fit range, we observe that our results are in perfect agreement with a linear behavior in a wide range of values of $n_\cool/\xi_L^2$, up to $n_\cool/\xi_L^2 \sim 14 $, see Fig.~\ref{fig:zerocool_limit_ex}. This is non-trivial, as this value corresponds to $n_\cool = 250$ and $n_\cool=690$ for, respectively, the smallest/largest $\xi_L$ investigated at this value of $N$. Reducing the fit range by choosing a smaller upper bound resulted in no relevant change in the obtained extrapolation. In any case, we incorporated any small variation observed into our final error, displayed as a full point and a shaded area in Fig.~\ref{fig:systematics_zerocool}.

As a final remark, we recall that, in order to correctly keep into account correlations among continuum extrapolations obtained for different values of $n_\cool/\xi_L^2$, we repeated the analysis described in this section for $5000$ different bootstrap resamples for each $\xi_L$, each one having the same size of the original sample. Thus, our results for the double extrapolation for different fit ranges, shown in Fig.~\ref{fig:systematics_zerocool}, were estimated by observing the variation of the double limit over such resamplings.

\subsection{The large-$N$ limit of $\chi^\prime$}

We repeat the analysis exemplified in the previous section for $N=5,11,31$. Obtained results are reported in Tab.~\ref{tab:chip_large_N_res}, while more details about the double extrapolation for these values of $N$ can be found in App.~\ref{app:other_double_extrapolations}.

\begin{table}[!t]
\begin{center}
\begin{tabular}{|c|c|}
\hline
$N$ & $\chi^\prime$ \\
\hline
5  &  0.0120(90)  \\
11 &  0.0032(15)  \\
21 &  -0.0020(35) \\
31 &  -0.0010(30) \\
\hline
\end{tabular}
\end{center}
\caption{Lattice determinations of $\chi^\prime$ as a function of $N$ after the double continuum $+$ zero-cooling extrapolation.}
\label{tab:chip_large_N_res}
\end{table}

\begin{figure}[!htb]
\centering
\includegraphics[scale=0.34]{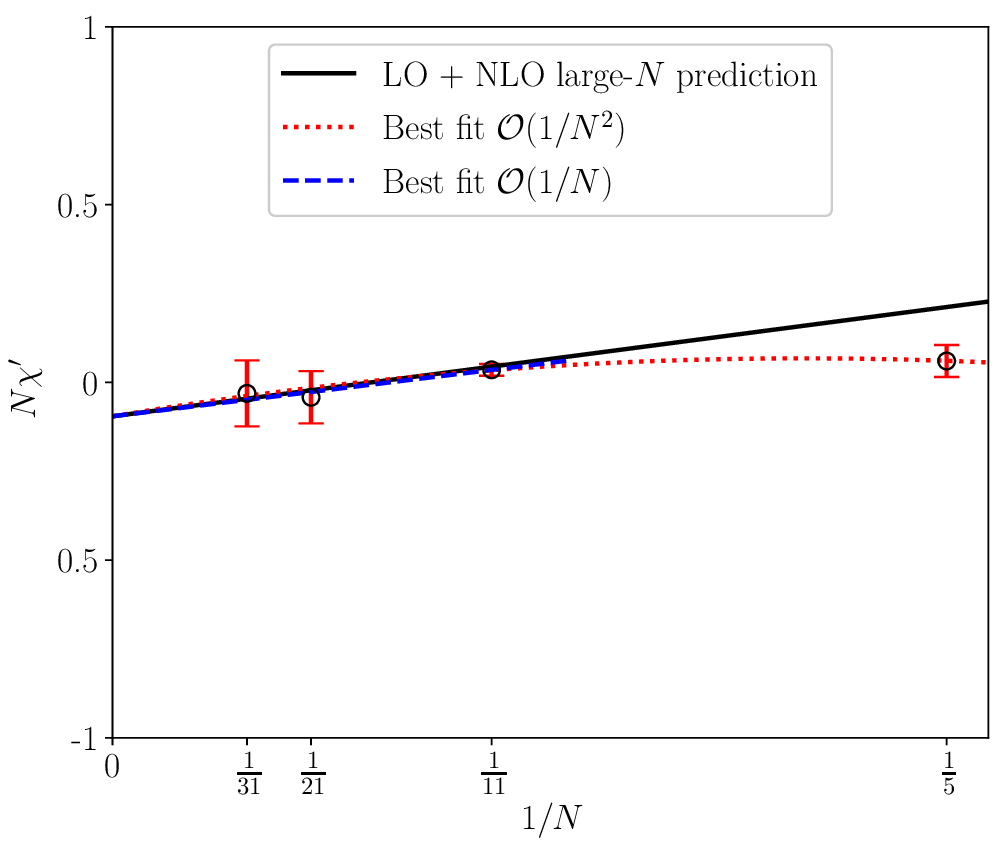}
\caption{Large-$N$ behavior of $\chi^\prime$ compared to the LO $+$ NLO analytic $1/N$ prediction. Dotted and dashed lines represent polynomial best fits of our lattice determinations of $\chi^\prime$.}
\label{fig:chip_largeN}
\end{figure}

Our aim is now to compare our results with the large-$N$ behavior predicted via the analytic $1/N$ expansion in Eq.~\eqref{eq:chip_largeN_analytic}. Such comparison is shown in Fig.~\ref{fig:chip_largeN}. Although our error bars are quite large (with our $N=21$ and $31$ determinations being compatible with 0 within errors), it is clear that our numerical results are in very good agreement with the large-$N$ analytic ones, and approach the NLO prediction already for $N\ge 11$.

If we try a best fit of our data for $N\chi^\prime$ with a polynomial in $1/N$, fixing the large-$N$ limit to $e_1^\prime=-3/(10\pi)$,
\beq
N \chi^\prime = e_1^\prime + e_2^\prime \frac{1}{N} + e_3^\prime \frac{1}{N^2},
\eeq
we find $e_2^\prime=1.44(18)$ (analytic: $1.53671$) by fitting up to $\mathcal{O}(1/N)$ terms in the range $N\ge 11$, and $e_3^\prime =-5.9(2.6)$ by fitting up to $\mathcal{O}(1/N^2)$ terms in the range $N\ge 5$ (with $e_2^\prime$ staying within errors). Leaving $e_1^\prime$ as a free parameters gives perfectly compatible results, but within larger error bars. These best fits are displayed in Fig.~\ref{fig:chip_largeN} as dashed/dotted lines.

Our result for $e_2^\prime$ is thus in perfect agreement with the large-$N$ analytic computation. We are also able to give a preliminary estimation of $e_3^\prime$, which turns out to be larger by about a factor of $4$ and of opposite sign compared to $e_2^\prime$. These findings match very well with those found for the large-$N$ behavior of the susceptibility $\chi$ and of the quartic coefficient $b_2$ in Ref.~\cite{Berni:2019bch}, where the coefficients of the $1/N$ expansion of these quantities appear to grow in absolute value with alternating signs as the order of $1/N$ is increased. This is not surprising, being the $1/N$ expansion an asymptotic series.

As a final remark, we stress that the difficulty in observing a clear signal above zero for $\chi^\prime$ for $N=21,31$ is due to the fact that $\chi^\prime$ is suppressed in the large-$N$ limit as $1/N$ and is very small already at small $N$, and is thus not related to any intrinsic drawback of any of the methods here employed. Therefore, this fact does not spoil the feasibility of our strategies in more complex models such as $4d$ $\SU(N)$ Yang--Mills theories, where such suppression at large-$N$ is not expected to occur.

\section{Conclusions}\label{sec:final}

In this paper we have presented a complete lattice investigation of the topological susceptibility slope $\chi^\prime$ in $2d$ $\CP^{N-1}$ models in the large-$N$ limit. Our main purpose was to test the feasibility of our numerical strategies in view of an application to $4d$ gauge theories, where this quantity plays an intriguing theoretical and phenomenological role.

Our approach relies on the parallel tempering on boundary conditions to mitigate topological freezing, and on a double extrapolation (continuum limit followed by the zero-cooling limit) to correctly reconstruct $\chi^\prime$. This strategy ensures that no relevant UV contribution is lost because of the smoothing method used to properly identify the topological charge of lattice field configurations.

We computed $\chi^\prime$ for $N=5,11,21$ and $31$ and found that our results are in very good agreement with those obtained from the large-$N$ analytic $1/N$ expansion. Therefore, this confirms the solidity and the feasibility of our numerical methods to compute $\chi^\prime$ from the lattice.

For this reason, in the near future we aim at employing such strategies also for $4d$ gauge theories. In particular, our next step will be to compute $\chi^\prime$ in the $4d$ $\SU(3)$ pure-gauge theory, in view of a lattice investigation of its large-$N$ limit in Yang--Mills theories, which has relevant theoretical interest in relation with the Witten--Veneziano mechanism. In the next future we also plan to perform an investigation of $\chi^\prime$ in full QCD, where its value has intriguing phenomenological implications.

\acknowledgments
I am deeply grateful to M.~D'Elia for many useful discussions and sincere encouragement, and for reading and sharing comments on this manuscript. It is also a pleasure to thank C.~Bonati for useful comments. I acknowledge the support of the Italian Ministry of Education, University and Research under the project PRIN 2017E44HRF, ``Low dimensional quantum systems: theory, experiments and simulations''. Numerical simulations have been performed on the \texttt{MARCONI} machine at CINECA, based on the agreement between INFN and CINECA, under projects INF21\_npqcd and INF22\_npqcd.

\appendix

\FloatBarrier

\section{Double extrapolation of $\chi^\prime$ for $N=5, 11, 31$}\label{app:other_double_extrapolations}

\begin{figure}
\centering
\includegraphics[scale=0.22]{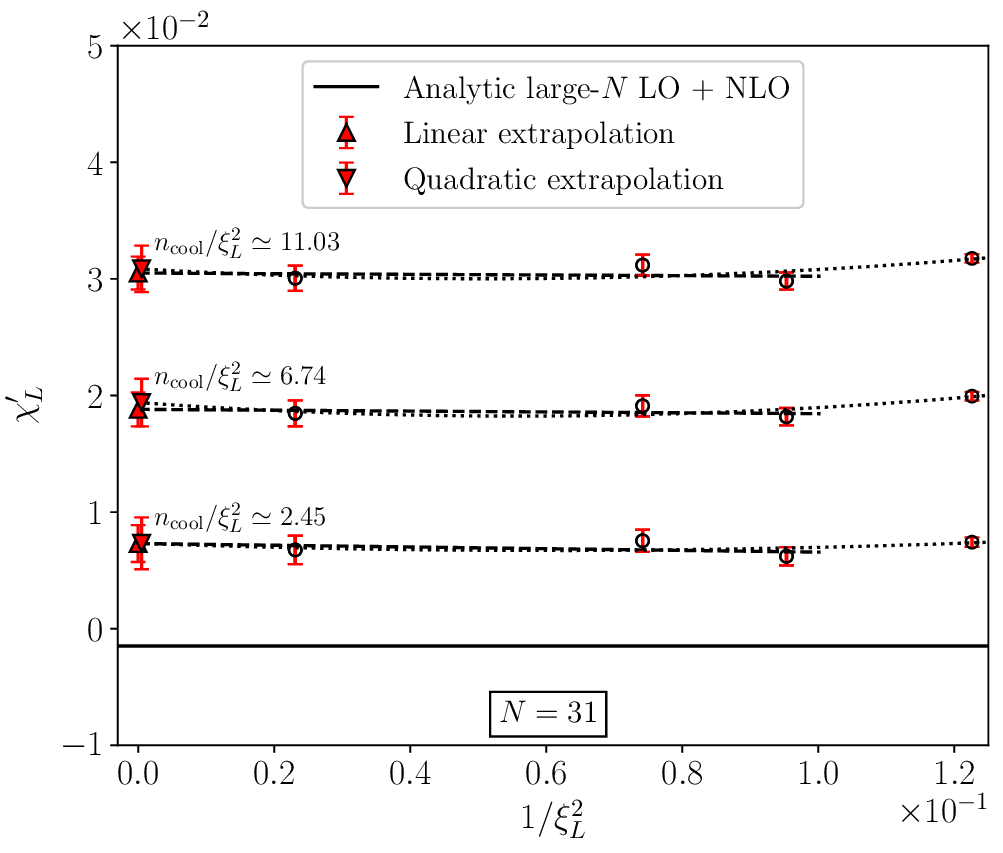}
\includegraphics[scale=0.22]{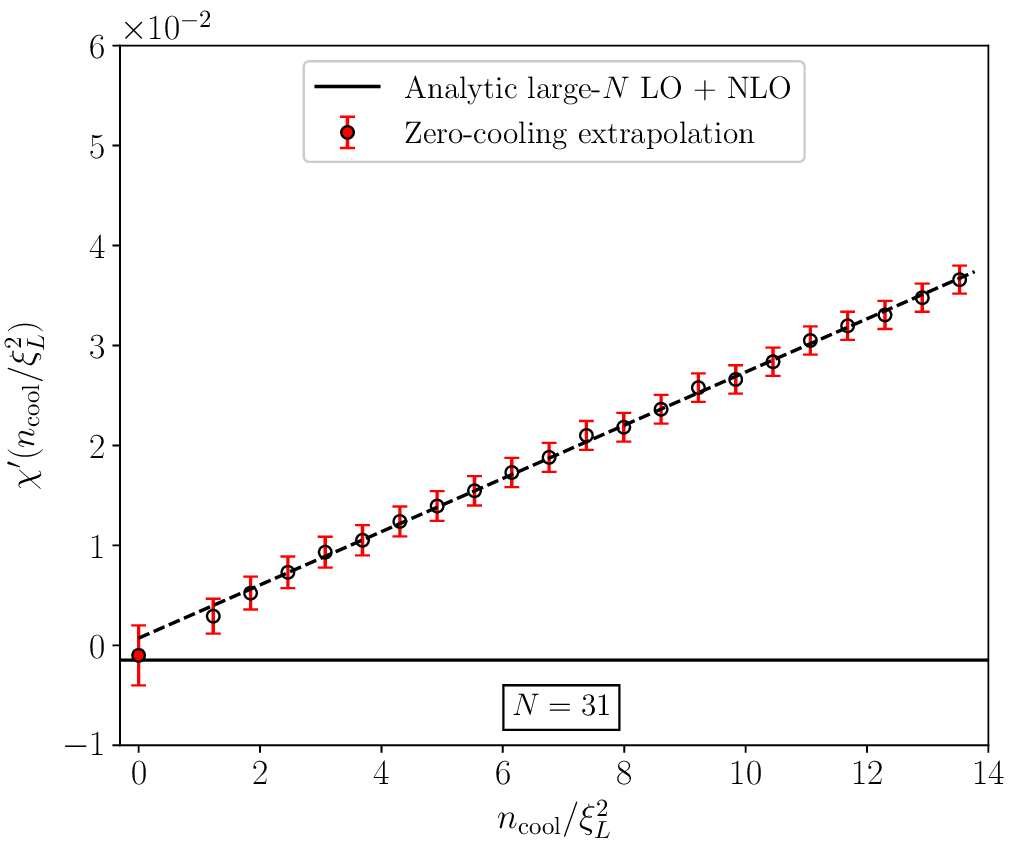}
\includegraphics[scale=0.22]{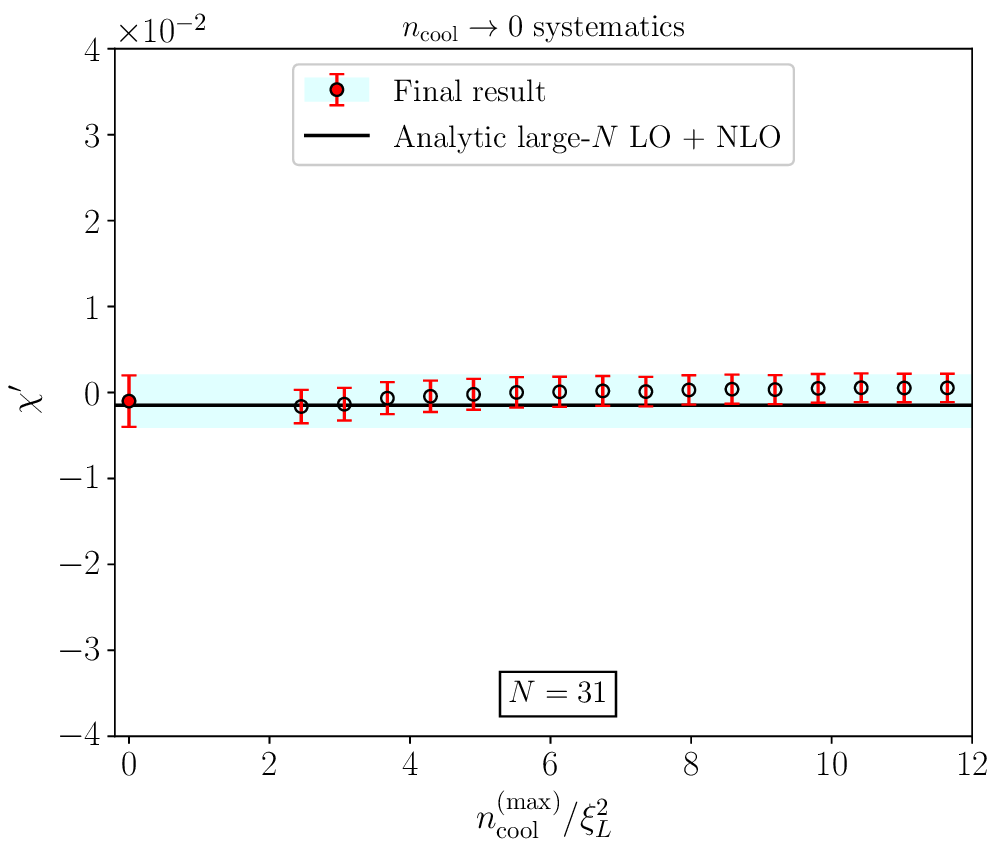}
\caption{Figures refer to $N=31$. Top left: examples of extrapolation towards the continuum limit of $\chi_L$ at fixed $n_{\cool}/\xi_L^2$. Top right: extrapolation towards the zero-cooling limit with a linear function in $n_\cool/\xi_L^2$. Bottom: results of the zero-cooling extrapolation varying the upper bound of the fit range $n_\cool^{(\max)}/\xi_L^2$. Final result for $\chi^\prime$ is shown as a full point in $n_\cool = 0$.}
\label{fig:res_N_31}
\end{figure}

\begin{figure}
\centering
\includegraphics[scale=0.22]{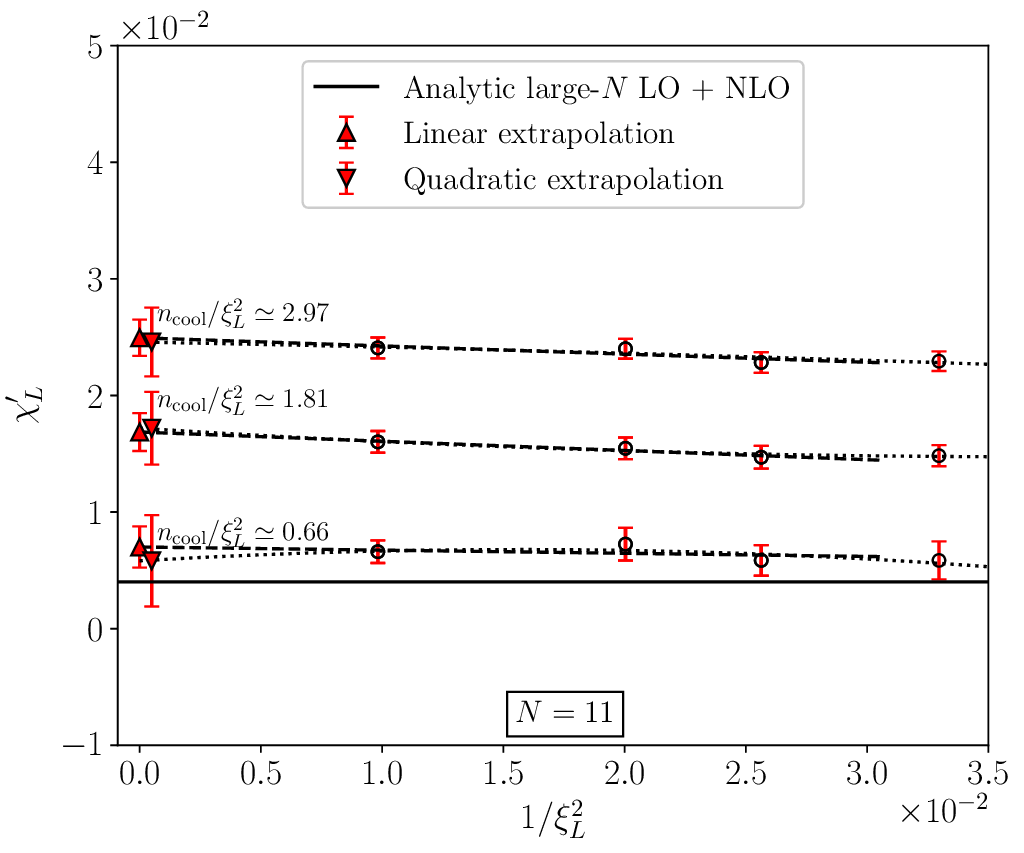}
\includegraphics[scale=0.22]{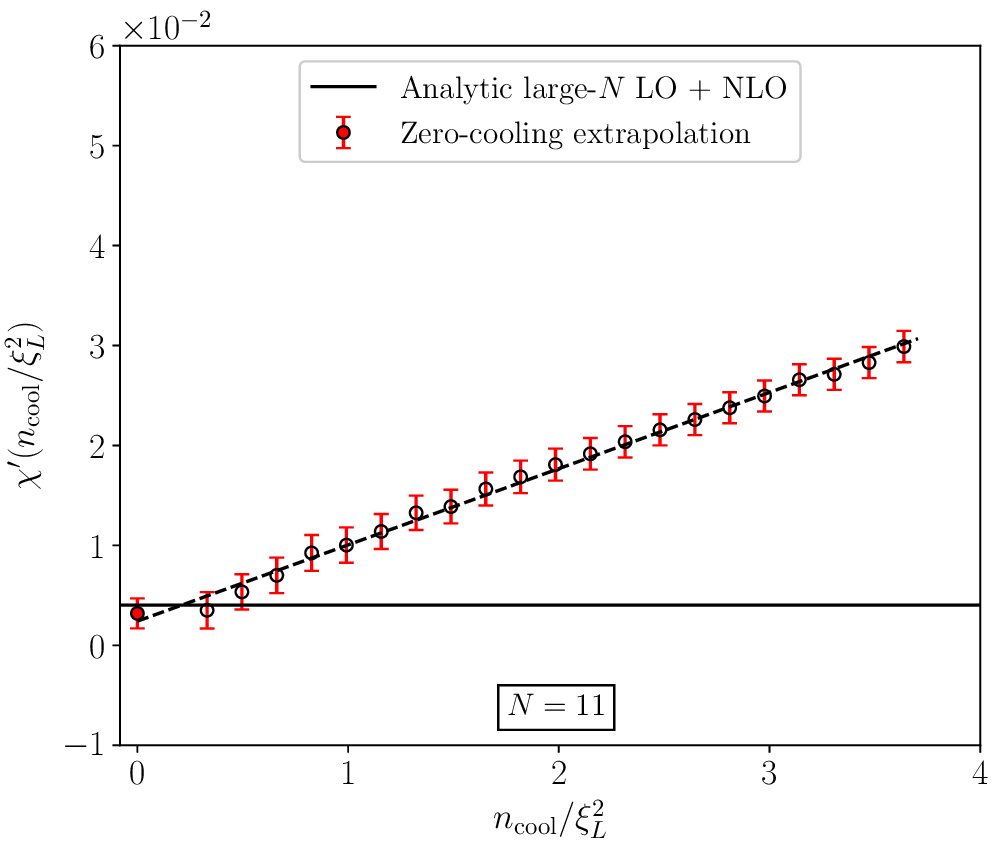}
\includegraphics[scale=0.22]{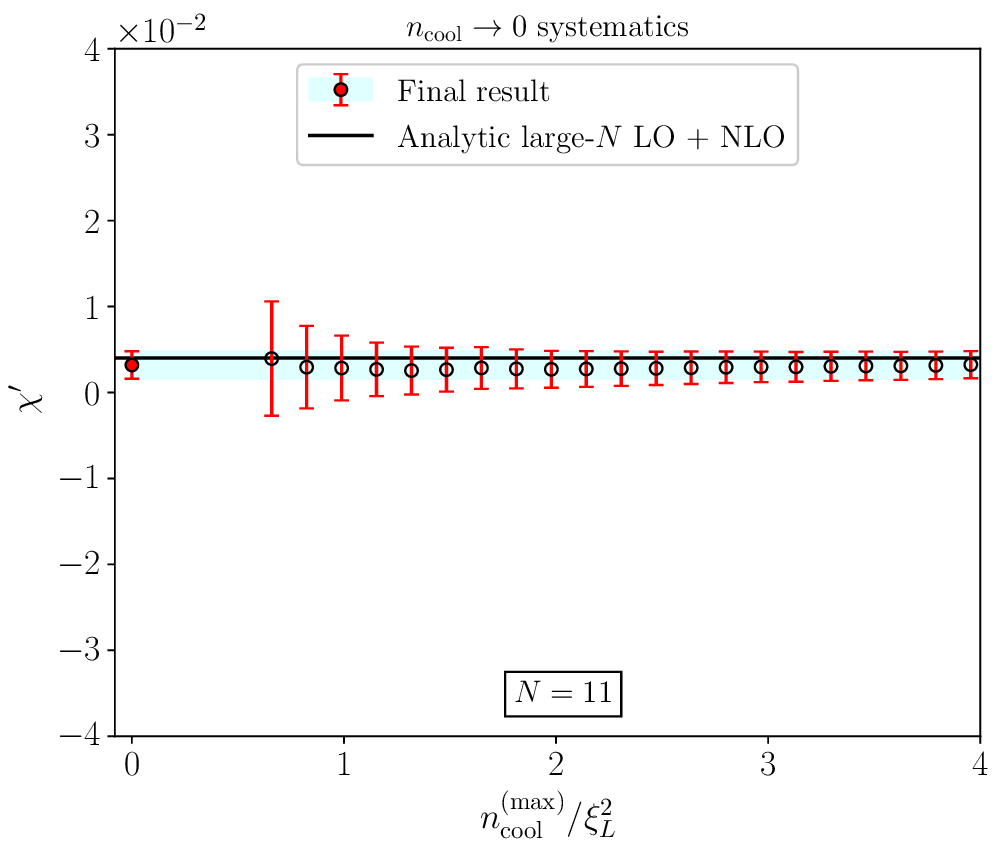}
\caption{Figures refer to $N=11$. Top left: examples of extrapolation towards the continuum limit of $\chi_L$ at fixed $n_{\cool}/\xi_L^2$. Top right: extrapolation towards the zero-cooling limit with a linear function in $n_\cool/\xi_L^2$. Bottom: results of the zero-cooling extrapolation varying the upper bound of the fit range $n_\cool^{(\max)}/\xi_L^2$. Final result for $\chi^\prime$ is shown as a full point in $n_\cool = 0$.}
\label{fig:res_N_11}
\end{figure}

\begin{figure}
\centering
\includegraphics[scale=0.22]{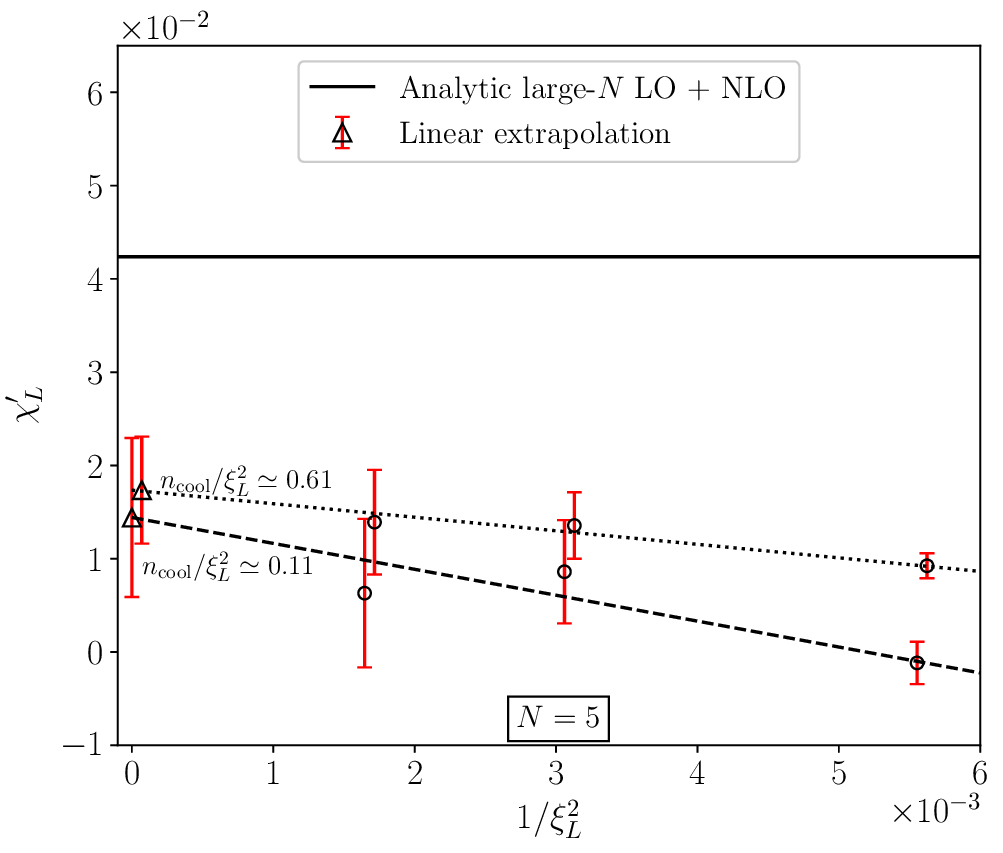}
\includegraphics[scale=0.22]{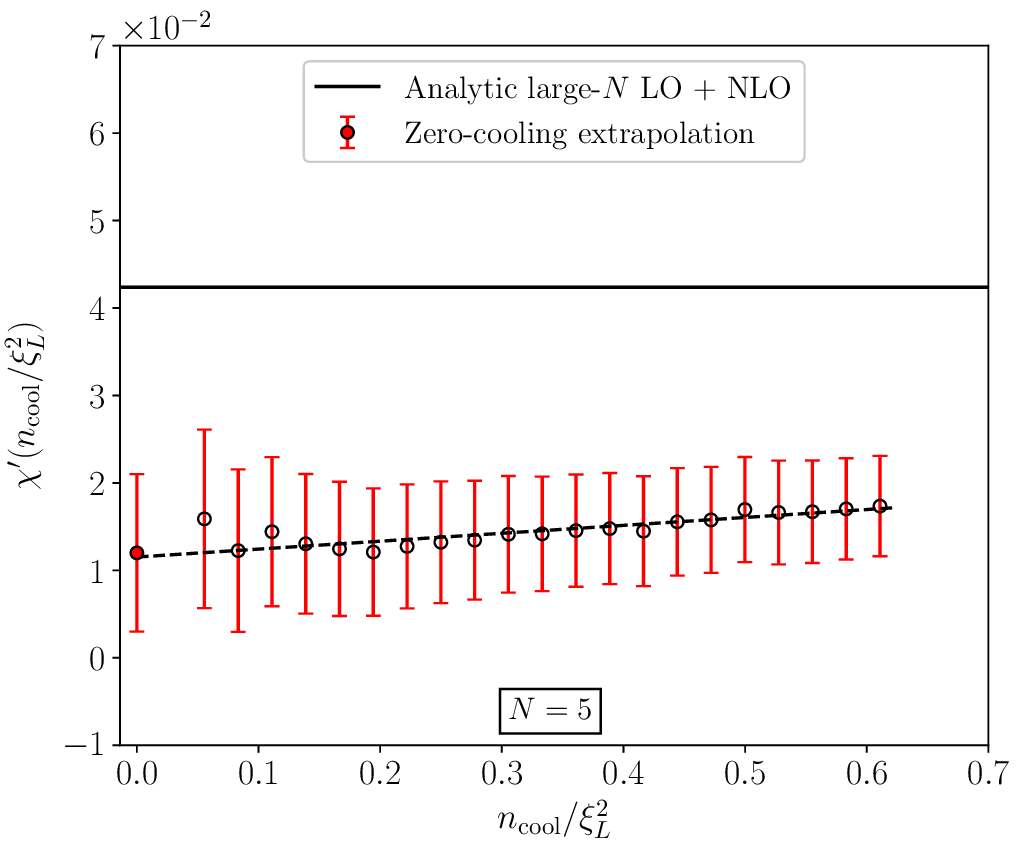}
\includegraphics[scale=0.22]{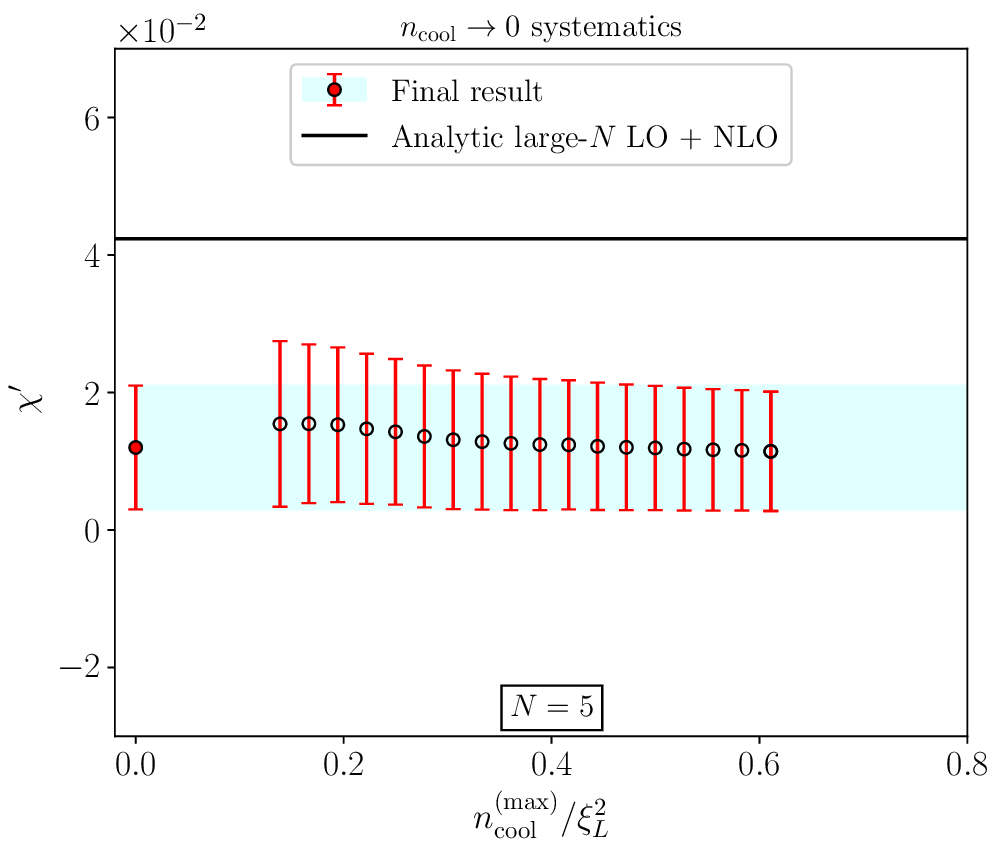}
\caption{Figures refer to $N=5$. Top left: examples of extrapolation towards the continuum limit of $\chi_L$ at fixed $n_{\cool}/\xi_L^2$. Top right: extrapolation towards the zero-cooling limit with a linear function in $n_\cool/\xi_L^2$. Bottom: results of the zero-cooling extrapolation varying the upper bound of the fit range $n_\cool^{(\max)}/\xi_L^2$. Final result for $\chi^\prime$ is shown as a full point in $n_\cool = 0$.}
\label{fig:res_N_5}
\end{figure}

In Figs.~\ref{fig:res_N_31},~\ref{fig:res_N_11} and~\ref{fig:res_N_5} we show the double extrapolation of $\chi^\prime$ for $N=5,11,31$ as done for $N=21$ in Sec.~\ref{sec:double_extrapolation_ex}. Results are displayed as follows: few examples of continuum extrapolation at fixed value of $n_\cool / \xi_L^2$, linear extrapolation in $n_\cool/\xi_L^2$ towards the zero-cooling limit, systematics related to the zero-cooling best fit.

\FloatBarrier

\end{document}